\documentclass[11pt,a4paper]{article}

\usepackage{mathtools}
\usepackage{authblk} 
\usepackage{algpseudocode,algorithmicx,algorithm}

\usepackage{mathrsfs}
\usepackage{latexsym,bm}
\usepackage{amsmath,amsfonts,amssymb,amsthm}
\usepackage{extarrows}

\usepackage{graphicx,subfigure,epstopdf,float}
\usepackage{enumerate,cases,multirow}
\usepackage{makecell}
\usepackage{caption}

\usepackage{longtable,colortbl,arydshln,threeparttable}
\definecolor{mygray}{gray}{.9}

\usepackage{indentfirst}
\usepackage[top=25mm,bottom=20mm,left=25mm,right=20mm]{geometry}
\usepackage{cite}

\usepackage{listings}

\usepackage{makeidx}        
\usepackage{booktabs}
\usepackage[bookmarksnumbered,colorlinks,citecolor=red,linkcolor=red,hyperindex,linktocpage=true]{hyperref}

\newcommand{\ket}[1]{| #1 \rangle} 
\newcommand{\bra}[1]{\langle #1 |} 

\newcommand{\bb}{\boldsymbol}

\def \d {\mathrm{d}}
\def \e {\mathrm{e}}
\def \i {\mathrm{i}}

\newcounter{parentalgorithm}

\makeatother

\newtheorem{theorem}{Theorem}[section]
\newtheorem{lemma}{Lemma}[section]

\theoremstyle{remark}
\newtheorem{remark}{\bf Remark}[section]

\numberwithin{equation}{section}

\begin{document}

\title{Investigation on a quantum algorithm for linear differential equations}

\author{Xiaojing Dong\thanks{dongxiaojing99@xtu.edu.cn}}
\author{Yizhe Peng\thanks{pengyizhe@smail.xtu.edu.cn} \footnote{Corresponding author.}}
\author{Qili Tang\thanks{tangqili@xtu.edu.cn}}
\author{Yin Yang\thanks{yangyinxtu@xtu.edu.cn}}
\author{Yue Yu\thanks{terenceyuyue@xtu.edu.cn}}
\affil{Hunan Key Laboratory for Computation and Simulation in Science and Engineering, Key Laboratory of Intelligent Computing and Information Processing of Ministry of Education, National Center for Applied Mathematics in Hunan, School of Mathematics and Computational Science, Xiangtan University, Xiangtan, Hunan 411105, PR China}

\maketitle

\begin{abstract}
  Ref.~\cite{BerryChilds2017ODE} introduced a pioneering quantum approach (coined BCOW algorithm) for solving linear differential equations with optimal error tolerance. Originally designed for a specific class of diagonalizable linear differential equations, the algorithm was extended by Krovi in \cite{KroviODE} to encompass broader classes, including non-diagonalizable and even singular matrices. Despite the common misconception, the original algorithm is indeed applicable to non-diagonalizable matrices, with diagonalisation primarily serving for theoretical analyses to establish bounds on condition number and solution error. By leveraging basic estimates from \cite{KroviODE}, we derive bounds comparable to those outlined in the Krovi algorithm, thereby reinstating the advantages of the BCOW approach. Furthermore, we extend the BCOW algorithm to address time-dependent linear differential equations by transforming non-autonomous systems into higher-dimensional autonomous ones, a technique also applicable for the Krovi algorithm.
\end{abstract}



\section{Introduction}

Quantum computing is an increasingly prominent computational paradigm that has garnered significant attention, primarily due to the discovery of quantum algorithms offering exponential acceleration over the best-known classical methods \cite{Nielsen2010, LR2010QuantumAlgebra, Deutsch1992rapid, Shor1997Prime, HHL2009}. In recent years, numerous quantum algorithms for scientific computing have been proposed. One fundamental problem in scientific computing is the development of solvers for linear systems of equations. In the realm of quantum computing, these are known as quantum linear systems algorithms (QLSAs), exemplified by the HHL algorithm \cite{HHL2009, Childs2017QLSA, Costa2021QLSA}. Proposed by Harrow, Hassidim, and Lloyd in 2009 \cite{HHL2009}, the HHL algorithm provides an early quantum approach to solving systems of linear equations, demonstrating exponential speedup over classical algorithms when the matrix's condition number is not excessively large.

Another critical challenge lies in developing solvers for linear ordinary differential equations (ODEs). The design of quantum algorithms for this task has been a subject of extensive discussion in the past few years and continues to draw significant attention today. The exponential speedup offered by the HHL algorithm holds promise for overcoming the curse of dimensionality, driving research into its application for solving linear systems arising from classical numerical discretisations of both ordinary and partial differential equations (PDEs). For instance, Berry \cite{Berry2014Highorder} first discretized first-order linear ODEs using linear multi-step methods and subsequently applied the HHL algorithm to solve the resulting linear systems. This approach, termed quantum difference methods in \cite{JLY2022multiscale}, has been further explored by Jin et al. who analyzed the time complexity of these methods, combining advanced QLSAs \cite{Costa2021QLSA} to address linear and nonlinear high-dimensional and multiscale PDEs within the framework of Asymptotic-Preserving schemes \cite{JLY22nonlinear,HJY2023Boltzmann}.

The HHL algorithm requires $\mathcal{O}(1/\delta)$ uses of a unitary operation to estimate eigenvalues with precision $\delta$. To mitigate the limitations of phase estimation, Childs et al. proposed an algorithm \cite{Childs2017QLSA} that exponentially improves dependence on the precision parameter,applying this method to develop finite difference and spectral algorithms for the Poisson equation and general second-order elliptic equations \cite{Childs2021high}. However, despite the improvements, the overall complexity remains $\text{poly}(1/\delta)$ when implementing the quantum difference method with multi-step methods \cite{Berry2014Highorder}.To overcome this, Berry et al. introduced a quantum algorithm for linear differential equations \cite{BerryChilds2017ODE} with complexity $\text{poly}(\log(1/\delta))$. This approach encodes a truncation of the Taylor series of the differential equation's propagator into a linear system, bypassing the use of multi-step methods. Initially developed for diagonalizable matrices, Krovi extended this algorithm \cite{KroviODE} to include non-diagonalizable matrices, demonstrating exponential speedup for certain classes compared to earlier bounds \cite{BerryChilds2017ODE}. Unlike sparse encoding methods \cite{BerryChilds2017ODE}, Krovi utilized block encoding due to involving inverses of small-scale matrices in the input model for the coefficient matrix. Further studies on the limitations and advancements of quantum algorithms for solving linear ODEs have been extensively covered in \cite{An2022blockEncodingODE}.

It is noteworthy that the BCOW algorithm introduced in \cite{BerryChilds2017ODE} in fact is applicable to non-diagonalizable matrices. Diagonalisation is primarily used for theoretical analysis to establish bounds on the condition number and solution error. By carefully comparing it with the Krovi algorithm in \cite{KroviODE}, we recognize that the enhanced complexity bound may be recovered for the BCOW algorithm. First, the Krovi algorithm involves the inversion of a small-scale matrix, with the inversion designed to eliminate intermediate variables in the linear system from \cite{BerryChilds2017ODE}. This elimination should not necessarily lead to a significant improvement in the condition number of the coefficient matrix. Second, the upper bound for the condition number of the matrix $ C_{m,k,p}(Ah)$ in \cite{BerryChilds2017ODE} is derived under the assumption that $ A $ is diagonalizable, specifically as $ A = VDV^{-1} $. Here, $ \kappa_V = \|V\| \|V^{-1}\| $ serves as a parameter in their analysis. However, as noted in the example in \cite{KroviODE}, this parameter can overestimate the condition number of $C_{m,k,p}(Ah)$. Third, the upper bound for the condition number of $C_{m,k,p}(Ah)$ is nearly comparable to that of the modified linear system for a special matrix $A$ discussed in Figure 2 of \cite{KroviODE}.  Conversely,  \cite{KroviODE} introduced a new parameter $ C(A)$, elaborated in Lemma \ref{Lem:boundary of T}, which offered a more precise upper bound for the condition number of the modified linear system in \cite{KroviODE}. Consequently, it is expected to enhance our understanding and analysis by characterizing the upper bound of the BCOW algorithm with this new parameter.

Building on the insights, this paper aims to establish bounds on the condition number and solution error using the parameter $C(A)$ introduced in \cite{KroviODE}.
With the assistance of basic estimates from \cite{KroviODE}, we successfully achieve these bounds with complexity comparable to that outlined in the Krovi algorithm.
Additionally, we extend the BCOW algorithm to address time-dependent linear differential equations. For a comprehensive review of related literature in this field, readers are referred to the recent paper \cite{BerryCosta2024timeODE}.
We remark that an innovative unitarisation approach known as the ``Schr\"odingerisation'' method has been introduced in \cite{JLY22SchrShort, JLY22SchrLong, analogPDE, CJL23TimeSchr} for both time-independent and time-dependent problems.
This method provides a straightforward and versatile framework enabling quantum simulation of all linear PDEs and ODEs. It employs a warped phase transform that maps equations into a higher dimension, where they manifest as systems of Schr\"odinger or Hamiltonian type equations in discrete Fourier space. A similar concept has also been explored in \cite{ALL2023LCH, ACL2023LCH2}, where Hamiltonian-type evolution is achieved in continuous Fourier space.
Drawing inspiration from \cite{CJL23TimeSchr}, we adapt techniques for transforming non-autonomous Hamiltonian systems into autonomous ones in a higher-dimensional context. This approach unifies the quantum simulation of non-autonomous systems with autonomous ones, avoiding cumbersome Dyson series expansions. The foundation for this technique can be traced back to \cite{Peskin19943tt}, where the time-dependent Schr\"odinger equation is transformed into a time-independent form using the $(t,t')$ method. In our work, we apply this transformative technique to non-Hamiltonian time-dependent problems and subsequently utilize the BCOW algorithm to analyze the resulting time-dependent system as well as provide detailed analyses of the time complexity in this context.

\section{The BCOW algorithm for linear differential equations}

In this section we review the quantum algorithm (referred to as the BCOW algorithm) in \cite{BerryChilds2017ODE} for solving the following system of linear ordinary differential equations
\begin{equation}\label{eq:ODEs}
	\begin{cases}
		\dfrac{\mathrm d\bb{x}(t)}{\mathrm dt}=A\bb{x}(t)+\bb{b}, \qquad t\in (0,T),\\
        \bb{x}(0)= \bb{x}_{in},
	\end{cases}
\end{equation}
where $A \in \mathbb{C}^{N\times N}$ and $\bb{b} \in \mathbb{C}^{N}$ are time independent. The BCOW algorithm is the first quantum ODE solver that exhibits the optimal dependence on the error tolerance. It should be pointed out that the time analysis there relies on the diagonalisation of the coefficient matrix $A$, however, the algorithm is valid for cases where $A$ is not diagonalisable.

The exact solution of \eqref{eq:ODEs} can be given in terms of the matrix exponential as
\begin{align*}
\bb{x}(t)
 = \e^{A t}\bb{x}(0) + \int_0^t \e^{A(t-s)}\d s \bb{b} =: T(At) \bb{x}(0) + S(At)\bb{b}, \\
\end{align*}
where
\[T(At) = \e^{A t}:= \sum\limits_{j=0}^\infty \frac{(At)^j}{j!}, \]
and
\[S(At) = \int_0^t \e^{A(t-s)}\d s = \int_0^t \e^{A s}\d s = (\e^{At}-I)(At)^{-1}t
= \sum\limits_{j=1}^\infty \frac{(At)^{j-1}}{j!} t .\]
Define
\begin{equation}\label{eq:T_k}
	T_k(Ah)=\sum_{j=0}^k\frac{(Ah)^j}{j!},
\end{equation}
and
\begin{equation}\label{eq:S_k}
	S_k(Ah)=\sum_{j=1}^k\frac{(Ah)^{j-1}}{j!} h.
\end{equation}
The solution can be approximated for a short evolution time $h$ by
\begin{equation}\label{eq:approximate}
	\bb{x}(h)\approx T_k(Ah)\bb{x}(0)+S_k(Ah)\bb{b},
\end{equation}
for sufficiently large truncation number $k$  (see Theorem \ref{thm:solutionerr} for the requirement of $h$). This approximate solution can be used in turn as an initial condition for another step of
evolution, and this procedure will be repeated for a total number of steps $N_t = m$ with the total evolution time $T = N_t h$.

According to \eqref{eq:T_k}-\eqref{eq:S_k}, the approximate solution \eqref{eq:approximate} can be expanded as
\begin{equation}\label{eq:iteration}
	\begin{aligned}
		\bb{x}(h)& \approx\Big(I+(Ah)+\frac{(Ah)^{2}}{2!}+\frac{(Ah)^{3}}{3!}+\cdots+\frac{(Ah)^{k}}{k!}\Big)\bb{x}(0)  \\
		&\quad+\left(I+\frac{(Ah)}{2!}+\frac{(Ah)^{2}}{3!}+\cdots+\frac{(Ah)^{k-1}}{k!}\right)(h\bb{b}) \\
		&=\bb{x}(0)+(Ah\bb{x}(0)+h\bb{b})+\frac{(Ah)}{2!}(Ah\bb{x}(0)+h\bb{b})+h\bb{b})+\cdots+\frac{(Ah)^{k-1}}{k!}(Ah\bb{x}(0)+h\bb{b}).
	\end{aligned}
\end{equation}
Following the process of converting higher-order linear PDEs into a system of first-order linear equations, we are ready to introduce the following variables
\begin{equation}
	\begin{cases}\label{eqs:F-O}
		\bb{x}_{0,0}=\bb{x}(0)=\bb{x}_{in},\\
		\bb{x}_{0,1}=(Ah)\bb{x}_{0,0}+(h\bb{b}),\\
		\bb{x}_{0,2}=(Ah/2)\bb{x}_{0,1},\\
		\bb{x}_{0,3}=(Ah/3)\bb{x}_{0,2},\\
		\vdots\\
		\bb{x}_{0,k}=(Ah/k)\bb{x}_{0,k-1},
	\end{cases}
\end{equation}
and obtain
\begin{equation}\label{eq:x_1}
	  \bb{x}(h)\approx \bb{x}_{1,0}:= \bb{x}_{0,0}+\bb{x}_{0,1}+\bb{x}_{0,2}+\bb{x}_{0,3}+\cdots+\bb{x}_{0,k}.
\end{equation}
The above system of linear equations \eqref{eqs:F-O}-\eqref{eq:x_1} is written in matrix form as
\begin{equation}\label{block1}
		\begin{bmatrix}
		I      &         &          &       &        &       \\
		-Ah    &    I    &          &       &        &       \\
		& -Ah/2   &    I     &       &        &       \\
		&         & \ddots   & \ddots &       &       \\
		&         &          & -Ah/k  &    I   &       \\
		-I     &   -I    &  \cdots  &   -I   &    -I   &    I
	\end{bmatrix}
	\begin{bmatrix}
		\bb{x}_{0,0} \\
		\bb{x}_{0,1} \\
		\bb{x}_{0,2} \\
		\vdots \\
		\bb{x}_{0,k} \\
		\bb{x}_{1,0}
	\end{bmatrix}=
	\begin{bmatrix}
		\bb{x}_{in} \\
		h\bb{b} \\
		\bb{0} \\
		\vdots\\
		\bb{0} \\
		\bb{0}
	\end{bmatrix}.
\end{equation}
Following the above procedure with $\bb{x}_{i-1,0}$ being the initial value, one has
\begin{equation}\label{block2}
		\begin{bmatrix}
		I      &         &          &       &        &       \\
		-Ah    &    I    &          &       &        &       \\
		& -Ah/2   &    I     &       &        &       \\
		&         & \ddots   & \ddots &       &       \\
		&         &          & -Ah/k  &    I   &       \\
		-I     &   -I    &  \cdots  &   -I   &    -I   &    I
	\end{bmatrix}
	\begin{bmatrix}
		\bb{x}_{i-1,0} \\
		\bb{x}_{i-1,1} \\
		\bb{x}_{i-1,2} \\
		\vdots \\
		\bb{x}_{i-1,k} \\
		\bb{x}_{i,0}
	\end{bmatrix}=
	\begin{bmatrix}
		\bb{x}_{i-1,0} \\
		h\bb{b} \\
		\bb{0} \\
		\vdots\\
		\bb{0} \\
		\bb{0}
	\end{bmatrix},
\end{equation}
for $i=2,\cdots,m$, where
\begin{equation}
	\bb{x}(ih) \approx \bb{x}_{i,0}: = \bb{x}_{i-1,0}+\bb{x}_{i-1,1}+\cdots+\bb{x}_{i-1,k}.
\end{equation}
To enhance the probability of achieving $\bb{x}(t=T) \approx \bb{x}_{m,0}$, several identities are typically appended at the end of the equation
\[
\bb{x}_{m,j}=\bb{x}_{m,j-1},\quad j=1,2,\cdots,p,
\]
which gives the last non-zero block

\begin{equation}\label{block3}
	\begin{bmatrix}
        I &\\
		-I & I \\
		& \ddots & \ddots\\
		& & -I & I\\
	\end{bmatrix}
	\begin{bmatrix}
        \bb{x}_{m,0}\\
		\bb{x}_{m,1}\\
		\vdots\\
		\bb{x}_{m,p}
	\end{bmatrix}
	=
	\begin{bmatrix}
        \bb{x}_{m,0}\\
		\bb{0} \\
		\vdots\\
		\bb{0}
	\end{bmatrix}.
\end{equation}

Let
\[\bb{X} = [\bb{x}_{0,1}; \cdots; \bb{x}_{0,k}; \bb{x}_{1,0};\cdots;\bb{x}_{1,k}; \cdots; \bb{x}_{m,0}; \bb{x}_{m,1}; \cdots; \bb{x}_{m,p}]\]
be the solution vector, with ``;'' indicating the straightening of $\{\bb{x}_{i,j}\}$ into a column vector. The number of vectors in $\{\bb{x}_{i,j}\}$ is denoted by $d+1$, where $d = m(k+1)+p$. Then the final system of linear equations can be formulated as
\begin{equation}\label{BCOWsystem}
C_{m,k,p}(Ah) \bb{X} = \bb{F},
\end{equation}
where
\[
\begin{aligned}
C_{m,k,p}(A)=\sum_{j=0}^{d}|j\rangle\langle j|\otimes I&-\sum_{i=0}^{m-1}\sum_{j=1}^{k}|i(k+1)+j\rangle\langle i(k+1)+j-1|\otimes A/j \\
&-\sum_{i=0}^{m-1}\sum_{j=0}^{k}|(i+1)(k+1)\rangle\langle i(k+1)+j|\otimes I-\sum_{j=d-p+1}^{d}|j\rangle\langle j-1|\otimes I,
\end{aligned}
\]
\[
 \bb{F}=|0\rangle \otimes \bb{x}_{in} + \sum_{i=0}^{m-1}|i(k+1)+1\rangle \otimes (h\bb{b}).
\]

For example, when $k=3$ and $m=p=2$, this gives
{\scriptsize\[
C_{m,k,p}(Ah) =
\begin{bmatrix}
I    &     &  & &  & & & & & &  \\
-Ah  &  I  &  & &  & & & & & &  \\
     & -Ah/2   &  I       &          & & & & & & & \\
     &         & -Ah/3    & I        & & & & & & & \\
-I   & -I      & -I       & -I       & I & & & & & & \\
 & & & &  -Ah & I & & & & & \\
 & & & & &       -Ah/2 & I     & & & & \\
 & & & &      &       & -Ah/3 & I & & & \\
 & & & &    -I &    -I &   -I  & -I & I & & \\
 & & & &       &       &       &    & -I & I & \\
 & & & &       &       &       &    &    & -I  & I
\end{bmatrix}, \]
\[\bb{X} = \begin{bmatrix} \bb{x}_{0,0}\\\bb{x}_{0,1}\\\bb{x}_{0,2}\\\bb{x}_{0,3} \\\bb{x}_{1,0} \\ \bb{x}_{1,1} \\ \bb{x}_{1,2} \\ \bb{x}_{1,3} \\ \bb{x}_{2,0}\\\bb{x}_{2,1}\\\bb{x}_{2,2} \end{bmatrix}, \qquad
\bb{F} = \begin{bmatrix} \bb{x}_{in} \\ h\bb{b}\\ \bb{0} \\ \bb{0} \\ \bb{0} \\ h\bb{b} \\ \bb{0} \\ \bb{0} \\\bb{0}\\\bb{0}\\\bb{0} \end{bmatrix}.\]
}

\begin{remark} \label{rem:TS}
It is apparent that
\[ T(Ah) = S(Ah)A + I, \qquad T_k(Ah) = S_k(Ah)A + I, \]
where we may leverage the commutativity of $A$ with its series.
\end{remark}

\section{New complexity bound for the BCOW algorithm}

\subsection{Solution error}

Let $\bb{x}_{i,h} = \bb{x}(ih)$ and $\bb{x}_{i,0}$ denote the exact and numerical solutions at $t_i = ih$, respectively, which satisfy
\begin{align}
& \bb{x}_{i+1,h} = T(Ah) \bb{x}_{i,h} + S(Ah) \bb{b} , \label{xih} \\
& \bb{x}_{i+1,0} = T_k(Ah) \bb{x}_{i,0} + S_k(Ah) \bb{b}. \label{xi0}
\end{align}
Let $t = T$. By recursively applying \eqref{xih}, we can derive
\begin{align*}
\bb{x}(t)
& = \bb{x}_{m,h} = T(Ah) \bb{x}_{m-1,h} + S(Ah) \bb{b} \\
& = T(Ah)\Big( T(Ah) \bb{x}_{m-2,h} + S(Ah) \bb{b} \Big)  + S(Ah) \bb{b} \\
& = T^2(Ah) \bb{x}_{m-2,h} + \sum_{j=0}^1 T^j(Ah) S(Ah) \bb{b}  \\
& = \cdots = T^m (Ah) \bb{x}_{in} + \sum_{j=0}^{m-1}T^j(Ah) S(Ah)\bb{b} =: \bb{x}^0(t) + \bb{x}^b(t).
\end{align*}
A direct manipulation gives
\[\bb{x}^0(t) = T^m (Ah) \bb{x}_{in} = \e^{At} \bb{x}_{in} = T(At) \bb{x}_{in},\]
and
\begin{align*}
\bb{x}^b(t)
& = \sum_{j=0}^{m-1}T^j(Ah) S(Ah)\bb{b}
  = \sum_{j=0}^{m-1} \e^{A jh} (\e^{A h} - I) (Ah)^{-1}h  \bb{b} \\
& = \sum_{j=0}^{m-1} \e^{A jh} \int_0^h \e^{A\tau} \d \tau \bb{b}
  = \sum_{j=0}^{m-1}  \int_0^h \e^{A(\tau+jh)} \d \tau \bb{b} \\
& \xlongequal{s = \tau+jh} \sum_{j=0}^{m-1}  \int_{jh}^{(j+1)h} \e^{A s } \d s \bb{b}
  = \int_0^t \e^{A s } \d s \bb{b} = S(At) \bb{b},
\end{align*}
implying that $\bb{x}^0(t)$ and $\bb{x}^b(t)$ are the exact solutions with homogeneous right-hand side (i.e., $\bb{b} = \bb{0}$) and zero initial data (i.e., $\bb{x}_{in} = \bb{0}$), respectively.
Accordingly, by recursively applying \eqref{xi0}, one can find that the approximate solution, denoted by $\tilde{\bb{x}}(t)$, can be written as
\begin{align*}
\tilde{\bb{x}}(t) = T_k^m (Ah) \bb{x}_{in} + \sum_{j=0}^{m-1}T_k^j(Ah) S_k(Ah)\bb{b} =: \tilde{\bb{x}}^0(t) + \tilde{\bb{x}}^b(t),
\end{align*}
with
\[\tilde{\bb{x}}^0(t) = T_k^m (Ah) \bb{x}_{in} =:\tilde{T}(A t)\bb{x}_{in} \]
and
\[\tilde{\bb{x}}^b(t) = \sum_{j=0}^{m-1}T_k^j(Ah) S_k(Ah)\bb{b} =:\tilde{S}(At)\bb{b} \]
being the approximate solutions of $\bb{x}^0(t)$ and $\bb{x}^b(t)$, respectively.

\begin{theorem} \label{thm:solutionerr}
Let $\ket{\bb{x}(T)}$ and $\ket{\tilde{\bb{x}}(T)}$ be the exact and approximate solution states, respectively. Suppose that the truncation number $k$ satisfies
\[(k+1)! \ge \frac{2 me^3}{\delta} \Big( 1 + \frac{T e^2\|\bb{b}\|}{\|\bb{x}(T)\|}\Big), \]
and the step size $h$ satisfies $\|Ah\| \le 1$. Then there holds
\[\|\ket{\bb{x}(T)} - \ket{\tilde{\bb{x}}(T)}\| \le \delta.\]
\end{theorem}
\begin{proof}
The proof is similar to that given in \cite[Theorem 3]{KroviODE}. Using the inequality $\| \bb{x}/{\|\bb{x}\|} - \bb{y}/{\|\bb{y}\|} \| \le 2 {\|\bb{x} - \bb{y}\|}/{\|\bb{x}\|}$ for two vectors $ \bb{x}, \bb{y}$, we can bound the error in the quantum state after a successful measurement as
\[
\| \ket{\bb{x}(t)} - \ket{\tilde{\bb{x}}(t)} \| \le \frac{2 \| \bb{x}(t) - \tilde{\bb{x}}(t) \|}{\| \bb{x}(t) \|} \le \delta, \qquad t = T,
\]
which gives
\[\| \bb{x}(t) - \tilde{\bb{x}}(t) \| \le \frac{\delta}{2}\| \bb{x}(t) \|.\]

According to the equalities in Remark \ref{rem:TS}, one can find that
\begin{align*}
\bb{x}(t)
= \bb{x}^0(t) + \bb{x}^b(t)
= T (At) \bb{x}_{in} + S(At)\bb{b}
= (S(At)A + I)\bb{x}_{in} + S(At)\bb{b},
\end{align*}
\[\tilde{\bb{x}}(t)
= \tilde{\bb{x}}^0(t) + \tilde{\bb{x}}^b(t)
= \tilde{T} (At) \bb{x}_{in} + \tilde{S} (At)\bb{b}
= (\tilde{S}(At)A + I) \bb{x}_{in} + \tilde{S}(At)\bb{b}. \]
For brevity, we omit ``$(At)$'' in the following.  By leveraging the commutativity of $A$ with its series, we obtain
\begin{align*}
\bb{x}(t) - \tilde{\bb{x}}(t)
& = (T-\tilde{T})  \bb{x}_{in} +  (S-\tilde{S}) \bb{b} \\
& = (T-\tilde{T}) T^{-1} ( \bb{x}(t) - S \bb{b}) +  (S-\tilde{S}) \bb{b} \\
& = (T-\tilde{T}) T^{-1} \bb{x}(t) + (SA-\tilde{S}A) T^{-1} ( - S\bb{b})) +  (S-\tilde{S}) \bb{b}\\
& = (T-\tilde{T}) T^{-1} \bb{x}(t) + (S-\tilde{S}) T^{-1} ( - AS + T  ) \bb{b}\\
& = (T-\tilde{T}) T^{-1} \bb{x}(t) + (S-\tilde{S}) T^{-1} \bb{b} =: I_1 + I_2,
\end{align*}
where
\[I_1 = (T-\tilde{T}) T^{-1}\bb{x}(t) = (T^m(Ah) - T_k^m(Ah) ) (T^m(Ah))^{-1}  \bb{x}(t),\]
\[I_2 = (S-\tilde{S}) T^{-1}  \bb{b} = \sum_{j=0}^{m-1}( T^j(Ah) S(Ah) - T_k^j(Ah) S_k(Ah) ) \bb{b}. \]

The estimates of $I_1$ and $I_2$ are shown in \cite{KroviODE} with the results given by
\begin{equation}\label{I1}
\|I_1\| \le \frac{me^3}{(k+1)!} \|\bb{x}(t)\|, \qquad \|I_2\| \le \frac{ t me^5}{(k+1)!} \|\bb{b}\|,
\end{equation}
where $k$ satisfies $me^2/(k+1)! \le 1$. Therefore,
\[\|\bb{x}(t) - \tilde{\bb{x}}(t) \| \le \frac{me^3}{(k+1)!} \Big( 1 + \frac{t e^2\|\bb{b}\|}{\|\bb{x}(t)\|}\Big)\|\bb{x}(t)\|
\qquad \mbox{if\quad $\dfrac{me^2}{(k+1)!} \le 1$}.\]
The proof is completed by requiring that the right-hand side is less than $\frac{\delta\| \bb{x}(t) \|}{2}$.
\end{proof}

With the estimate of $I_1$ in \eqref{I1}, we are ready to derive the bound for $\|T^\ell_k(Ah)\|$ for any $\ell \le m$.
\begin{lemma}\cite[Lemma 13]{KroviODE}\label{Lem:boundary of T}
Under the condition of Theorem \ref{thm:solutionerr}, we have that the truncated Taylor series $T_{k}( Ah)$ of $\e^{Ah}$ satisfies
	\[\|T^\ell_k(Ah)\|\leq C(A)(1+\delta),\]
	for any $\ell \leq m$, where
    \begin{equation}\label{CA}
    C(A):=\sup\limits_{t\in[0,T]} \|\e^{At} \|.
    \end{equation}
\end{lemma}

\subsection{Condition number}

In deriving an upper bound for the condition number of the matrix $ C_{m,k,p}(Ah) $, the authors of \cite{BerryChilds2017ODE} assume the matrix $ A $ to be diagonalizable, specifically as $ A = VDV^{-1} $. Here, the condition number $ \kappa_V = \|V\| \|V^{-1}\| $ serves as a parameter in their analysis. However, as noted in the example provided in \cite{KroviODE}, this parameter may significantly overestimate the condition number of $ C_{m,k,p}(Ah) $. On the other hand, the upper bound for the condition number of $ C_{m,k,p}(Ah) $ is nearly comparable to that of the modified linear system discussed in \cite{KroviODE}, where a more suitable parameter $ C(A) $ is introduced as given in Lemma \ref{Lem:boundary of T}. Consequently, it is expected to characterize the upper bound of the BCOW algorithm by using this new parameter.

The upper bound of the norm of $C_{m,k,p}(Ah)$ is provided in Lemma 4 of \cite{BerryChilds2017ODE} described as follows.

\begin{lemma}\label{lem:nm}	
Let $A$ be an $N \times N$ matrix and the step size $h$ satisfies $\| A h\| \le 1$. Let $m,k,p \in \mathbb{Z} ^+$, and the truncation number $k \ge 5$. Then
\[\|C_{m,k,p}(Ah )\| \le 2\sqrt{k}.\]
\end{lemma}

For the upper bound of $\|C_{m,k,p}(Ah)^{-1}\|$, we define
\[T_{b,k}(z) =\sum_{j=b}^k \frac{b!z^{j-b}}{j!},\]
where $b$ is an integer with $0\le b \le k$.  With the step size $h$ satisfying $\| Ah\| \le 1$, it is simple to find that
\begin{equation}\label{Tbk}
\|T_{b,k}(Ah)\| \le \sum_{j=0}^{k-b} \frac{1}{j!} \le e.
\end{equation}

\begin{theorem}\label{thm:kc}
Suppose that the truncation number $k \ge 5$. Under the condition of Theorem \ref{thm:solutionerr}, there holds
	\begin{equation}
		\kappa_C \leq 9 k(m+p)C(A)(1+\delta),
	\end{equation}
where $\kappa_C$ is the condition number of $C_{m,k,p}(Ah)$ and $C(A)$ is define in Lemma \ref{Lem:boundary of T}.
\end{theorem}
\begin{proof}
Following the approach in \cite{BerryChilds2017ODE,KroviODE}, we establish an upper bound using the definition
\[\|C^{-1}_{m,k,p}(Ah)\| = \sup_{\|\bb{B}\| \leq 1}\|C^{-1}_{m,k,p}(Ah)\bb{B}\|,\]
where $\bb{B}$ is a column vector.

The vector $\bb{B}$ will be partitioned according to the solution vector as
\[
\bb{B} = [\bb{\beta}_{0,0}; \cdots; \bb{\beta}_{0,k};\cdots;\bb{\beta}_{m-1,0};\cdots;\bb{\beta}_{m-1,k};\bb{\beta}_{m,0} ;\cdots ;\bb{\beta}_{m,p}].
\]
For simplicity, we relabel the vectors with a single index $ g $ as follows:
\[\bb{B} = \sum\limits_{g = 0}^d \ket{g} \otimes \bb{\beta}_{g} =: \sum\limits_{g = 0}^d  \bb{b}_{g}  , \]
where $d = m(k+1) + p$, $\ket{g}$ is the standard computational basis vector for $g = 0,1,\cdots,d$, and $\bb{\beta}_{g} = \bb{\beta}_{ij}$ with $g = i(k+1) + j$. Here, for $0 \le i < m$, $0 \le j \le k$ and for $i = m$, $0 \le j \le p$.

For every $\bb{b}_g$, we define
\begin{equation}\label{solvebg}
\bb{y}^g = C^{-1}_{m,k,p}(Ah)\bb{b}_g, \qquad g = 0,1,\cdots, d.
\end{equation}
It suffices to estimate the upper bound of $\|\bb{y}^g\|$.

Let $d_1 = m(k+1)$. We consider the cases $0 \le g \le d_1 -1$ and $n_1 \le g \le d$ separately, where $g$ represents the position of $\bb{\beta}_g$.

\textbf{Case 1}:  When $0 \leq g \leq d_1 - 1$, $\bb{\beta}_g$ corresponds to the blocks in  \eqref{block1} or \eqref{block2}. Specifically, there exist $0 \leq a < m$ and $0 \leq b \leq k$ such that
    \begin{equation}\label{a+1_block}
        \begin{bmatrix}
        I    &            &             &      &        &      &  \\
        -Ah  &    I       &             &      &        &      &  \\
             & \ddots     &  \ddots     &      &        &      &  \\
        	 &            &   -Ah/b     &  I   &        &      &   \\
             &            &             &  \ddots    & \ddots &      &  \\
             &            &             &      &  -Ah/k      &   I  &  \\
        -I   &   -I       & \cdots      &  -I  &   \cdots     &  -I    & I \\
		\end{bmatrix}
		\begin{bmatrix}
			\bb{x}^g_{a,0} \\
            \bb{x}^g_{a,1} \\
			\vdots \\
			\bb{x}^g_{a,b} \\
			\vdots \\
			\bb{x}^g_{a,k} \\
			\bb{x}^g_{a+1,0}
		\end{bmatrix}=
		\begin{bmatrix}
			\bb{0} \\
            \bb{0} \\
			\vdots \\
			\bb{\beta}_g \\
			\vdots\\
			\bb{0} \\
			\bb{0}
		\end{bmatrix}.
	\end{equation}
The other blocks of \eqref{solvebg} have the zero vector as their right-hand side.
Note that the matrices $(-I)$s in the first row are dropped since the vectors preceding $\bb{x}_{a,0}$ are zero. On the other hand, we assume that $\bb{\beta}_g$ is not located in the last row of the above block, otherwise we can put it in the next block.

We now present the explicit solution of \eqref{solvebg}. For the vectors preceding $\bb{x}_{a,b}^g$, one easily finds that
\[
\bb{x}_{i,j}^g = \bb{0}, \qquad 0 \le i < a, ~0 \le j \le k \quad \mbox{and} \quad i = a, ~0 \le j < b,
\]
which gives
\[\bb{x}^g_{a,b} = \bb{\beta}_g\]
and
\begin{align*}
& \bb{x}_{a,j}^g =  b!(Ah)^{j-b}/j! \bb{\beta}_g, \qquad b< j  \le k, \\
& \bb{x}_{a+1,0}^g = T_{b,k}(Ah)\bb{\beta}_g.
\end{align*}
For $i\ge a+1$, since the right-hand side is a zero vector, we obtain
\begin{align*}
& \bb{x}_{a+1,j}^g =((Ah)^{j}/j!) \bb{x}_{a+1,0}^g,  \qquad 0 \leq j\leq k, \\
& \bb{x}_{a+2,0}^g = T_{k}(Ah) \bb{x}_{a+1,0}^g = T_{k}(Ah)T_{b,k}(Ah)\bb{\beta}_g.
\end{align*}
Proceeding with this procedure, one can show that
\begin{align*}
& \bb{x}_{a+2,j}^g  = ((Ah)^{j}/j!) \bb{x}_{a+2,0}^g, \\
& \bb{x}_{a+3,0}^g = T_{k}(Ah) \bb{x}_{a+2,0}^g = (T_{k}(Ah))^2T_{b,k}(Ah)\bb{\beta}_g, \\
& \vdots \\
& \bb{x}_{m,0}^g = T_{k}(Ah) \bb{x}_{m-1,0}= (T_{k}(Ah))^{m-a-1}T_{b,k}(Ah)\bb{\beta}_g.
\end{align*}
Furthermore, the appended block \eqref{block3} gives
\[\bb{x}_{m,j}^g = \bb{x}^g_{m,0} = (T_{k}(Ah))^{m-a-1}T_{b,k}(Ah)\bb{\beta}_g, \qquad 1\le j \le p.\]
Since $\|A h\| \le 1$, for any $b\le j \le k$, we have
\[
\|\bb{x}_{a,j}^g\| \le  \|b!(Ah)^{j-b}/j! \bb{\beta}_g\| \le b!/j! \|\bb{\beta}_g\| = b!/j! \|\bb{b}_g\|.
\]
By using the estimates in Lemma \ref{Lem:boundary of T} and Eq.~\eqref{Tbk}, one can derive
\begin{align*}
\|\bb{x}_{i,0}^g\|
& \le \|T_{k}(Ah))^{i-a-1}\| \cdot \| T_{b,k}(Ah)\| \cdot \| \bb{\beta}_g\| \\
& \le  C(A)(1+\delta) e \| \bb{\beta}_g\| =  C(A)(1+\delta) e \| \bb{b}_g\|, \qquad a+1 \le i \le m,
\end{align*}
and
\[
\|\bb{x}_{i,j}^g \| \le \| ((Ah)^{j}/j!) \bb{x}_{i,0}^g\| \le C(A)(1+\delta) e/j! \| \bb{b}_g\|, \qquad a+1\le i \le m, \quad 1\le j \le k.
\]
Using these facts, we obtain
\begin{align*}
\|\bb{y}^g\|^2
& = \sum_{i=0}^{m-1} \sum_{j=0}^k  \| \bb{x}_{i,j}^g \|^2 + \sum_{j=0}^p  \| \bb{x}_{m,j}^g \|^2 \\
& = \sum_{j=b}^k ({b!}/{j!})^2 \|\bb{b}_g\|^2  + \sum_{i=a+1}^{m-1} \sum_{j=0}^k (C(A)(1+\delta) e/j!)^2 \| \bb{b}_g\|^2
  + (p+1) \|\bb{x}_{m,0}\|^2 \\
& \le I_0(2) \|\bb{b}_g\|^2 + (C(A)(1+\delta) e )^2(m-a-1) \sum_{j=0}^k (1/j!)^2 \| \bb{b}_g\|^2 + (p+1) ( C(A)(1+\delta) e)^2  \| \bb{b}_g\|^2 \\
& \le I_0(2) \|\bb{b}_g\|^2 + (C(A)(1+\delta) e )^2(m-a-1) I_0(2) \| \bb{b}_g\|^2 + (p+1) ( C(A)(1+\delta) e)^2  \| \bb{b}_g\|^2 \\
& \le I_0(2)(C(A)(1+\delta) e )^2(m+p) \| \bb{b}_g\|^2,
\end{align*}
where we have used the facts (see Eqs.~(50) and (51) in \cite{BerryChilds2017ODE})
\begin{align*}
& \sum_{j=0}^k (1/j!)^2 \le \sum_{j=0}^\infty (1/j!)^2 =: I_0(2)<2.28, \\
& \sum_{j=b}^k ({b!}/{j!})^2 < I_0(2).
\end{align*}

\textbf{Case 2}: When $d_1 \le  g \le d$,  $\beta_g$ corresponds to the block in \eqref{block3}, which can be written as
\[
    \begin{bmatrix}
    I    &            &             &      &        &      &  \\
    -I  &    I       &             &      &        &      &  \\
         & \ddots     &  \ddots     &      &        &      &  \\
    	 &            &   -I     &  I   &        &      &   \\
         &            &             &  \ddots    & \ddots &      &  \\
         &            &             &      &  -I      &   I  &  \\
       &          &       &    &        &  -I    & I \\
    		\end{bmatrix}
    		\begin{bmatrix}
    			\bb{x}^g_{m,0} \\
                \bb{x}^g_{m,1} \\
    			\vdots \\
    			\bb{x}^g_{m,b} \\
    			\vdots \\
    			\bb{x}^g_{m,p-1} \\
    			\bb{x}^g_{m,p}
    		\end{bmatrix}=
    		\begin{bmatrix}
    			\bb{0} \\
                \bb{0} \\
    			\vdots \\
    			\bb{\beta}_g \\
    			\vdots\\
    			\bb{0} \\
    			\bb{0}
    		\end{bmatrix},
\]
where $0 \leq b \leq p$. In this case, we get
\begin{align*}
&\bb{x}_{i,j}^g = \bb{0},\qquad  & 0\le i < m, \quad 0\le j\le k,\\
&\bb{x}_{m,j}^g = \bb{0},\qquad & 0\le j<b,\\
&\bb{x}_{m,j}^g = \bb{\beta} _g ,\qquad & b\leq j\leq p,
\end{align*}
yielding
\[
\|\bb{y}^g\|^2 = (p-b+1)\|\bb{\beta}_g\|^2 \le  (p+1) \| \bb{b}_g \|^2.
\]
Therefore, for any $0 \le g \le n-1$, there holds
\[
\|\bb{y}^g\|^2 = \|C^{-1}_{m,k,p}\bb{b}^g \|^2 \le I_0(2)(C(A)(1+\delta) e )^2(m+p) \| \bb{b}_g\|^2,
\]
leading to
\begin{align*}
\|C^{-1}_{m,k,p}(Ah)\|^2
\le (d+1) I_0(2)(C(A)(1+\delta) e )^2(m+p), \qquad d = m(k+1) + p
\end{align*}
or
\begin{align*}
\|C^{-1}_{m,k,p}(Ah)\|
& \le \sqrt{I_0(2) (m(k+1) + p + 1) (m+p)} C(A)(1+\delta) e \\
& \le \sqrt{\frac{6I_0(2)}{5}}  e \sqrt{ k } (m+p) C(A) (1+\delta) \\
& \le 4.5\sqrt{ k } (m+p) C(A) (1+\delta),
\end{align*}
where we have used
\begin{align}
(m(k+1)+p+1)(m+p)
& = (mk+1)(m+p)+(m+p)^2 \nonumber \\
&\leq (mk+kp)(m+p)+(m+p)^2 \nonumber\\
&=(k+1)(m+p)^2
\leq \frac{6}{5}k(m+p)^2, \qquad k \ge 5. \label{inversebound}
\end{align}

Combining the upper bounds in \eqref{inversebound} and Lemma \ref{lem:nm}, we obtain
\[
\kappa_C \leq 9k(m+p)C(A)(1+\delta).
\]
This completes the proof.
\end{proof}

\subsection{New complexity bound}

The complexity is quantified in terms of the following oracles (see Lemma 8 in \cite{BerryChilds2017ODE}):
\begin{itemize}
  \item The input model for the coefficient matrix $A$: an oracle $O_A$ that computes the non-zero entries of $A$.

  \item The source term input oracle $O_b$:  a controlled oracle that prepares the state proportional to $\bb{b}$.

  \item The state preparation oracle $O_x$: a controlled oracles that prepares the state proportional to $\bb{x}_{in}$.
\end{itemize}

\begin{theorem}\label{thm:complexity}
Suppose that $A$ is an $N\times N$ matrix with sparsity $s$ and $C(A)$ defined in \eqref{CA}. Let $\bb{x}(t)$ evolve according to the differential equation \eqref{eq:ODEs}. Let $T>0$ and $g:= \max_{t\in [0,T]}\frac{\|\bb{x}(t)\|}{\|\bb{x}(T)\|}$.
Then there exists a quantum algorithm that produces a state $\epsilon$-close to $\bb{x}(T)/\|\bb{x}(T)\|$ in $\ell^2$ norm, succeeding with probability $\Omega(1)$, with a flag indicating success, using
\[\mathcal{O}\Big( C(A) s T \|A\| \cdot \mathrm{poly} \Big( \log \frac{C(A) s T \|A\|g \beta   }{\epsilon}  \Big) \Big)\]
queries to $O_A$, $O_x$ and $O_b$, where $\beta = 1 + \frac{T e^2\|\bb{b}\|}{\|\bb{x}(T)\|}$.
\end{theorem}

\begin{proof}
We use the quantum linear system algorithm (QLSA) from \cite{Childs2017QLSA} to solve \eqref{BCOWsystem} and obtain a state $\ket{\bb{x}'}$ such that $\| \ket{\bb{x}} - \ket{\bb{x}'}\| \le \delta$. According to Theorem 5 of \cite{Childs2017QLSA}, the QLSA for $Ax = b$ makes $\mathcal{O}(s_A\kappa_A  \text{poly} ( \log (s_A \kappa_A/\delta) ) )$ queries to oracles for a $s_A$-sparse matrix $A$ and for preparing a state proportional to $b$ are needed to produce a state proportional to $A^{-1}b$ up to error $\delta$.

First, we consider building the linear system $C_{m,k,p}(Ah) \bb{X} = \bb{F}$ in \eqref{BCOWsystem}.
According to Lemma 8 in \cite{BerryChilds2017ODE}, the state preparation oracle for $\bb{F}$ can be produced with a constant number of calls to $O_x$ and $O_b$. Similarly, the matrix $C_{m,k,p}(Ah)$ can be constructed with a constant number of calls to $O_A$.

Second, we pick the parameters following \cite{BerryChilds2017ODE}:
\[h = \frac{T}{\lceil T \|A\|\rceil}, \qquad m = p = \frac{T}{h} = \lceil T \|A\|\rceil, \qquad \delta = \frac{\epsilon}{25\sqrt{m} g}, \qquad \epsilon<\frac12,\]
\[k = \Big\lceil \frac{2 \log \Omega}{\log \log \Omega} \Big\rceil,\]
where
\[\Omega = \frac{2 me^3}{\delta} \Big( 1 + \frac{T e^2\|\bb{b}\|}{\|\bb{x}(T)\|}\Big)\]
and $\delta$ is the solution error given in Theorem \ref{thm:solutionerr}. With these choices, we can find (see the proof of Theorem 9 of \cite{BerryChilds2017ODE}):
\begin{itemize}
  \item $\|Ah \| \le 1$ and  $(k+1)!\ge \Omega$;

  \item the QLSA outputs the state $\ket{\bb{X}'}$ which is $\delta$-close to $\bb{\ket{X}}$;

  \item the measurement produces an state $\ket{\bb{x}'(T)}$ which is $\epsilon$-close to $\ket{\bb{x}(T)}$ with probability $\gtrsim 1/g^2$.
\end{itemize}
The success probability can be raised to $\Omega(1)$ by using $\mathcal{O}(g)$ rounds of amplitude amplification.

The final step is to compute the complexity. The matrix $C_{m,k,p}(A)$ has $\mathcal{O}(ks)$ nonzero entries in any row or column. By Theorem \ref{thm:kc} and our choice of parameters, the condition number of $C_{m,k,p}(A)$ is
\[\mathcal{O}(  k(m+p)C(A)(1+\delta) ) = \mathcal{O}(  k T \|A\| C(A) ) .\]
Consequently, by Theorem 5 of \cite{Childs2017QLSA}, the QLSA produces the state produces state $\ket{\bb{x}'}$ with
\begin{align*}
 \mathcal{O}\Big( s k^2 T \|A\| C(A)\text{poly} \Big( \log \frac{s k^2 T\|A\| C(A)}{\delta}   \Big) \Big)
 = \mathcal{O}\Big( C(A) s T \|A\| \text{poly} \Big( \log \frac{C(A) s T \|A\|g \beta   }{\epsilon}  \Big) \Big)
\end{align*}
queries to $O_A$, $O_x$ and $O_b$, where $\beta = 1 + \frac{T e^2\|\bb{b}\|}{\|\bb{x}(T)\|}$. Since amplitude amplification requires $\mathcal{O}(g)$ repetitions of this procedure, the final query complexity will be multiplied by $g$, as claimed.
\end{proof}

\section{BCOW algorithm for time-dependent linear differential equations}

This section extends the BCOW algorithm to the following time-dependent problem
\begin{equation}\label{Timeu}
\frac{\d }{\d t}\bb{x}(t) = A(t)\bb{x}(t) + \bb{b}(t), \qquad \bb{x}(0) =  \bb{x}_0,
\end{equation}
 where both the coefficient matrix $ A $ and the right-hand side $ \boldsymbol{b} $  vary with time. The extension employs the autonomisation method introduced in \cite{CJL23TimeSchr}, which transforms a non-autonomous system into an autonomous one in one higher dimension.

\subsection{The autonomisation method}

\subsubsection{An enlarged system}

We first employ the augmentation technique in \cite{JLY22SchrLong,JLLY2024boundary,JLM24SchrInhom,JLY24Circuits} to transform equation \eqref{Timeu} to the homogeneous case. Denote $F(t) = \text{diag}(\bb{f}(t))$ as a diagonal matrix with the $i$-th entry of $\bb{f}(t)$ given by
 \[f_i(t) = \frac{b_i(t)}{( (b_i^2)_{ave} + \varepsilon^2 )^{1/2}}, \qquad i=0,1,\cdots,N-1,\]
 where
 \[
 (b_i^2)_{ave} := \frac{1}{T} \int_0^T |b_i(t)|^2 \d t
 \]
 and  $\varepsilon = 1/\sqrt{N}$ is included to prevent division by zero in the denominator.
 Then one can rewrite \eqref{eq:ODEs} as an enlarged system
\begin{equation}\label{enlargeTime}
\frac{\d }{\d t} \bb{u}(t)
= \begin{bmatrix}
A(t)  &  F(t) \\
O     &  O
\end{bmatrix}\bb{u}(t)=: B(t)\bb{u}(t), \qquad \bb{u}(t) = \begin{bmatrix}
\bb{x}(t) \\
\bb{r}(t)
\end{bmatrix}  , \qquad
\bb{u}(0) = \begin{bmatrix}
\bb{x}(0) \\
\bb{r}(0)
\end{bmatrix},
\end{equation}
	where $\bb{r}(t)$ is a constant column vector with the $i$-th entry given by
	\[r_i(t) = ( (b_i^2)_{ave} + \varepsilon^2 )^{1/2}, \qquad i=0,1,\cdots, N-1.\]
	For this new system, the probability of projecting onto $\ket{\bb{x}(t)}$ is
	\begin{equation}\label{Prenlarge}
		\text{Pr} = \frac{\|\bb{x}(t)\|^2}{\|\bb{u}(t)\|^2}
		= \frac{\|\bb{x}(t)\|^2}{\|\bb{x}(t)\|^2 + \|\bb{b}\|_{ave}^2+1},
	\end{equation}
with $\|\bb{b}\|_{ave} = \Big(\sum\limits_{i=0}^{N-1} (b_i^2)_{ave}\Big)^{1/2}$.

	The solution to \eqref{enlargeTime} with initial data at $t = t_0$ defines a time evolution operator, $\mathcal{U}_{t,t_0}$, satisfying
	\[\bb{u}(t) = \mathcal{U}_{t,t_0} \bb{u}_0, \qquad \bb{u}_0 = \bb{u}(t_0),\]
	where the propagator can be expressed as a time-ordered exponential \cite{Dyson1949Series,Dyson1949Smatrix,Low2019Interaction,Berry2019Dyson,BerryChilds2020TimeHamiltonian,CJL23TimeSchr}):
\[\mathcal{U}_{t,t_0} = \exp_{\mathcal{T}}\Big(\int_{t_0}^t B(\tau) \d \tau \Big)
=: I + \sum_{n=1}^\infty \frac{1}{n!} \int_{t_0}^t \d t_1  \int_{t_0}^t \d t_2 \cdots \int_{t_0}^t \d t_n \mathcal{T}[B(t_1)B(t_2)\cdots B(t_n)],\]
where the time-ordering operator $\mathcal{T}$ sorts any sequence of $n$ operators according to the times $t_j$ of their evaluation, that is,
\[\mathcal{T}[B(t_1)B(t_2)\cdots B(t_n)] = B(t_{i_1})B(t_{i_2})\cdots B(t_{i_n}), \qquad t_{i_1}> t_{i_2} > \cdots > t_{i_n}.\]

\begin{lemma}\cite{CJL23TimeSchr,Peskin19943tt} \label{lem:autonomisationTheorem}
For the non-autonomous system in \eqref{enlargeTime},  introduce the following initial-value problem of an autonomous PDE
\begin{align*}
& \frac{\partial \bb{z}}{\partial t} + \frac{\partial \bb{z}}{\partial s} = B(s) \bb{z}(t,s),\\
& \bb{z}(0,s) = G(s)\bb{u}_0, \qquad s\in [-T,T],
\end{align*}
where $G(s)$ is a single-variable function and $H(s) = 0$ if $s<0$.
The analytical solution to this problem is
\[\bb{z}(t,s) = G(s-t)\mathcal{U}_{s,s-t}\bb{u}_0,\]
where
\[
\mathcal{U}_{s,s-t} = \exp_{\mathcal{T}}\Big( \int_{s-t}^s B(\tau) \d \tau \Big) = \exp_{\mathcal{T}}\Big( \int_0^t B(s-t+\tau) \d \tau \Big).
\]
The solution to \eqref{enlargeTime} can be expressed in terms of $\bb{u}$ as
\begin{equation}\label{retriveu}
\bb{z}(t,s = t) = G(0) \bb{u}(t), \qquad t\ge t_0.
\end{equation}
\end{lemma}
\begin{proof}
This lemma is demonstrated in \cite{CJL23TimeSchr} for Hamiltonian systems by verifying the given result. Here we provide an alternative proof for non-unitary dynamics. Define $\bb{v}(t; s') = \bb{z}(t, s'+t)$, where $s'\in \mathbb{R}$ is a parameter. It is straightforward to show that $\bb{v}(t; s')$ satisfies
 \begin{align*}
& \frac{\d \bb{v}(t; s')}{\d t}  = \frac{\partial \bb{z}}{\partial t}(t, s'+t) + \frac{\partial \bb{z}}{\partial s}(t, s'+t) =  B(s'+t) \bb{v}(t'; s'),\\
& \bb{v}(0; s') = G(s')\bb{u}_0.
\end{align*}
The solution can be expressed using the time-ordered exponential as
\[\bb{v}(t; s') = \exp_{\mathcal{T}}\Big( \int_0^t B(s'+\tau) \d \tau \Big)G(s')\bb{u}_0. \]
Setting $s' = s-t$, we immediately obtain
\[\bb{z}(t,s) = \bb{v}(t; s-t) = \exp_{\mathcal{T}}\Big( \int_0^t B(s-t+\tau) \d \tau \Big)G(s-t)\bb{u}_0,\]
which completes the proof.
\end{proof}

To apply the discrete Fourier transform to $s$, it is necessary to require that $\bb{z}(t,s)$ is periodic in the $s$ direction. To this end, we choose $G(s)$ as the mollifier defined by
\[G(s) = c \eta(s), \qquad  \eta(x) = \begin{cases}
\e^{1/(x^2-1)}, \qquad & |x|<1, \\
0, \qquad & \mbox{otherwise},
\end{cases}\]
where $c = e$ gives $G(0) = 1$ and the support of $G$ is always in $[-1,1]$. According to Lemma \ref{lem:autonomisationTheorem}, this choice imposes the periodic boundary condition
\[\bb{z}(t,s=-1+T) = \bb{z}(t, s = 1+T) = \bb{0}\]
in the $s$ direction.

\subsubsection{Discretisation of the autonomisation problem}

We choose a uniform mesh size $\Delta s = 2/N_s$ for the autonomisation variable with $N_s = 2^{n_s}$ being an even number,
with the grid points denoted by $-1+T =  s_0<s_1<\cdots<s_{N_s} = 1+T$. We place the $s$-register in the first position and define
\[\bb{w}(t) = \sum_{l=0}^{N_s-1}\sum_{i=0}^{2N-1} \bb{z}_i(t,s_l) \ket{l,i},\]
where $\bb{z}_i$ is the $i$-th entry of $\bb{z}$.
By applying the discrete Fourier transformation in the $s$  direction, one arrives at
\begin{equation}\label{autodiscretisation}
\frac{\d}{\d t} \bb{w}(t) = \bar{A} \bb{w}(t), \qquad \bb{w}(0) = [G(s_0),\cdots, G(s_{N_s-1})]^T\otimes \bb{u}(0).
\end{equation}
Here the evolution matrix is
\begin{equation}\label{autoA}
\bar{A} = -\i P_s \otimes I^{\otimes (n+1)} + \sum_{l=0}^{N_s-1} \ket{l}\bra{l} \otimes B(s_l), \qquad N = 2^n,
\end{equation}
with
\[P_s = F_s D_s F_s^{-1},  \qquad D_s = \text{diag}(\mu_0, \cdots, \mu_{N_s-1}),  \qquad \mu_l = \pi \Big(l - \frac{N_s}{2}\Big),\]
where $F_s$ is the matrix representation of the discrete Fourier transform.

\begin{remark}\label{rem:Ns}
It is evident that $\|D_s\|_{\max} \le \frac{\pi N_s}{2}$.  Noting that the solution in the $s$ direction is smooth and the Fourier spectral method has spectral accuracy, we can choose $N_s = \mathcal{O}( \log (1/\delta) )$ with $\delta$ being the error bound.
\end{remark}

\subsection{BCOW algorithm for solving the autonomisation problem}

Applying the BCOW algorithm to \eqref{autodiscretisation} yields a system similar to \eqref{BCOWsystem}:
\[C(\bar{A} h) \bar{\bb{X}} = \bar{\bb{F}},\]
where $\bar{\bb{F}}=|0\rangle \otimes \bb{w}(0)$ and
\[
\begin{aligned}
C_{m,k,p}(\bar{A})=\sum_{j=0}^{d}|j\rangle\langle j|\otimes I&-\sum_{i=0}^{m-1}\sum_{j=1}^{k}|i(k+1)+j\rangle\langle i(k+1)+j-1|\otimes \bar{A}/j \\
&-\sum_{i=0}^{m-1}\sum_{j=0}^{k}|(i+1)(k+1)\rangle\langle i(k+1)+j|\otimes I-\sum_{j=d-p+1}^{d}|j\rangle\langle j-1|\otimes I.
\end{aligned}
\]

The matrix $\bar{A}$ may be constructed from the sparse encoding of $D_s$ (and two applications of the quantum Fourier transform) and $B(s_l)$. For sparse encoding of time-dependent matrices, we refer the reader to \cite{Low2019Interaction} for example.  The initial data $\bb{w}(0)$ can be obtained from the preparation oracles for $[G(s_0), \ldots, G(s_{N_s-1})]^T$ and $\bb{u}(0)$. For simplicity, we assume access only to the oracle $O_{\bar{A}}$, which computes the non-zero entries of $\bar{A}$, and the state preparation oracle $O_w$, responsible for preparing the state proportional to $\bb{w}(0)$.

By leveraging the findings from Theorem \ref{thm:complexity}, we are in a position to establish the following main result.

\begin{theorem}
Suppose that $A(t)$ is an $N\times N$ matrix with sparsity $s$. Let $\bb{x}(t)$ evolve according to the differential equation \eqref{Timeu}. Let $T>0$. In addition, assume that $\bb{x}^{(l)}(t)$ is the solution of
\begin{equation*}
\begin{cases}
\dfrac{\d \bb{x}^{(l)}(t)}{\d s} = A(s_l)\bb{x}^{(l)}(t) + \bb{b}(s_l), \\
\bb{x}^{(l)}(0) = G(s_l)\bb{x}(0), \qquad l = 0,1,\cdots, N_s-1.
\end{cases}
\end{equation*}
Then there exists a quantum algorithm that produces a state $\delta$-close to $\bb{x}(T)/\|\bb{x}(T)\|$ in $\ell^2$ norm, succeeding with probability $\Omega(1)$, with a flag indicating success, using
\[\mathcal{O}\Big( s C_A \|A\|_m T \mathrm{poly}\Big(\log \frac{s C_A \|A\|_m T g }{\delta}\Big) \Big)\]
queries to $O_{\bar{A}} $ and $O_w$, where
\[C_A =  \max_l C(A(s_l)), \qquad \|A\|_m =  1 + \max_l \|A(s_l)\| + \max_{i,l} |f_i(s_l)|, \]
\[g = \frac{\max_{l,t} \|\bb{x}^{(l)}(t)\| + \|\bb{b}\|_{ave}+ 1 }{\|\bb{x}(t)\|} \max_l ( 1 + T \|\bb{f} (s_l)\|), \]
and
\[f_i(t) = \frac{b_i(t)}{( (b_i^2)_{ave} + \varepsilon^2 )^{1/2}}, \qquad i=0,1,\cdots,N-1,\]
\[\|\bb{b}\|_{ave} = \Big(\sum\limits_{i=0}^{N-1} (b_i^2)_{ave}\Big)^{1/2}, \qquad (b_i^2)_{ave} := \frac{1}{T} \int_0^T |b_i(t)|^2 \d t.\]
\end{theorem}

\begin{proof}
(1) We retrieve the solution following this procedure:
\[
\ket{\bar{\bb{X}}} \rightarrow \ket{\bb{w}(t)} \rightarrow \ket{\bb{u}(t)} \rightarrow \ket{\bb{x}(t)}.
\]
First, we solve the linear system for the BCOW algorithm to obtain the state vector $\ket{\bar{\bb{X}}}$, which can be projected onto $\ket{\bb{w}(t)}$ with a probability given by $\frac{\|\bb{w}(t)\|^2}{\max_{t} \|\bb{w}(t)\|^2}$. Next, we derive $\ket{\bb{u}(t)}$ from $\ket{\bb{w}(t)}$ according to \eqref{retriveu} with a probability given by $\frac{\|\bb{u}(t)\|^2}{\|\bb{w}(t)\|^2}$. Finally, the desired state $\ket{\bb{x}(t)}$ is obtained from the enlarged variable in \eqref{enlargeTime} with probability $\frac{\|\bb{x}(t)\|^2}{\|\bb{u}(t)\|^2}$. The overall probability is
\[
\frac{\|\bb{w}(t)\|^2}{\max_{t} \|\bb{w}(t)\|^2}  \frac{\|\bb{u}(t)\|^2}{\|\bb{w}(t)\|^2}  \frac{\|\bb{x}(t)\|^2}{\|\bb{u}(t)\|^2} = \frac{\|\bb{x}(t)\|^2}{\max_{t} \|\bb{w}(t)\|^2}.
\]
By applying amplitude amplification as described in \cite{BerryChilds2017ODE}, we can increase this probability to $\Omega(1)$ with $\mathcal{O}(g_w)$ repetitions of the above procedure, where $g_w := \frac{\max_{t} \|\bb{w}(t)\|}{\|\bb{x}(t)\|}$.

In the following, we replace $\max_{t} \|\bb{w}(t)\|$ by quantities related to $\bb{x}(t)$. Noting that $\|\bb{w}\|^2 = \bb{w}^\dag \bb{w}$, $\bb{w}^\dag \frac{\d \bb{w}}{\d t}  = \bb{w}^\dag \bar{A} \bb{w}$ and $\frac{\d \bb{w}^\dag}{\d t} \bb{w} = \bb{w}^\dag \bar{A}^\dag \bb{w}$, we obtain
\[\frac{\d \|\bb{w}\|^2}{\d t} = \bb{w}^\dag (\bar{A}+ \bar{A}^\dag) \bb{w}
= \bb{w}^\dag\Big( \sum_{l=0}^{N_s-1} \ket{l}\bra{l} \otimes (B(s_l) + B^\dag(s_l)) \Big)\bb{w},\]
where the imaginary part or the first term on the right-hand side of \eqref{autoA} is naturally eliminated since $P_s$ is a Hermitian matrix. Therefore, when analyzing the evolution of the norm of $\bb{w}$, we can disregard the imaginary term in \eqref{autoA}. In other words, $\bb{w}(t)$ can be assumed to be the solution of \eqref{autodiscretisation} with the first term on the right-hand side omitted. To distinguish, we denote the corresponding variable as $\tilde{\bb{w}}(t)$, which satisfies $\frac{\d \|\bb{w}\|^2}{\d t} = \frac{\d \|\tilde{\bb{w}}\|^2}{\d t}$ with the same initial condition.
We can rewrite $\tilde{\bb{w}}$ as
\begin{equation}\label{tildewblock}
\tilde{\bb{w}}(t) =: [\bb{u}^{(0)}(t); \bb{u}^{(1)}(t); \cdots ;\bb{u}^{(N_s-1)}(t)],
\end{equation}
with ``;'' indicating the straightening of $\{\bm{u}^{(i)}\}_{i\ge 1}$ into a column vector. Since $\sum_{l=0}^{N_s-1} \ket{l}\bra{l} \otimes B(s_l)$ is a block diagonal matrix, we have
\[\frac{\d}{\d t} \bb{u}^{(l)}(t) = B(s_l)\bb{u}^{(l)}(t), \qquad \bb{u}^{(l)}(0) = G(s_l)\bb{u}(0), \qquad l = 0,1,\cdots,N_s-1,\]
which can be rewritten as
\begin{equation*}
\frac{\d }{\d t} \bb{u}^{(l)}(t)
= \begin{bmatrix}
A(s_l)  &  F(s_l) \\
O     &  O
\end{bmatrix}\bb{u}^{(l)}(t), \qquad \bb{u}^{(l)}(t) = \begin{bmatrix}
\bb{x}^{(l)}(t) \\
\bb{r}^{(l)}(t)
\end{bmatrix}  , \qquad
\bb{u}^{(l)}(0) = G(s_l)\begin{bmatrix}
\bb{x}(0) \\
\bb{r}(0)
\end{bmatrix}.
\end{equation*}
Here $\bb{r}^{(l)}(t) = G(s_l) \bb{r}$ is a constant vector and $\bb{x}^{(l)}(t)$ satisfies
\begin{equation} \label{xlt}
\begin{cases}
\dfrac{\d \bb{x}^{(l)}(t)}{\d s} = A(s_l)\bb{x}^{(l)}(t) + \bb{b}(s_l), \\
\bb{x}^{(l)}(0) = G(s_l)\bb{x}(0), \qquad l = 0,1,\cdots, N_s-1.
\end{cases}
\end{equation}
Therefore,
\[\frac{\d}{\d t}\|\bb{w}(t)\|^2 = \frac{\d}{\d t}\|\tilde{\bb{w}}(t)\|^2 = \sum_{l=0}^{N_s-1} \frac{\d}{\d t}\|\bb{x}^{(l)}(t)\|^2\]
with each $\bb{x}^{(l)}(t)$ satisfying \eqref{xlt}. This gives
\[\|\bb{w}(t)\|^2 = \|\bb{w}(0)\|^2 + \sum_{l=0}^{N_s-1} ( \|\bb{x}^{(l)}(t)\|^2 - \|\bb{x}^{(l)}(0)\|^2) \]
and
\begin{align*}
\max_{t\in [0,T]} \|\bb{w}(t)\|^2
& =  (\|\bb{x}(0)\|^2 + \|\bb{b}\|_{ave}^2+1) \sum_{l=0}^{N_s-1} |G_{s_l}|^2  + \max_{t\in [0,T]} \sum_{l=0}^{N_s-1} (\|\bb{x}^{(l)}(t)\|^2 - \|\bb{x}^{(l)}(0)\|^2) \\
& = (\|\bb{b}\|_{ave}^2+1) \sum_{l=0}^{N_s-1} |G_{s_l}|^2  + \max_{t\in [0,T]} \sum_{l=0}^{N_s-1} \|\bb{x}^{(l)}(t)\|^2 .
\end{align*}
According to Remark \ref{rem:Ns}, we have
\begin{align*}
\max_{t\in [0,T]} \|\bb{w}(t)\| \le  N_s^{1/2} \Big( \|\bb{b}\|_{ave}^2+ 1 + \max_l \max_{t\in [0,T]} \|\bb{x}^{(l)}(t)\|^2 \Big)^{1/2},
\end{align*}
where $N_s = \mathcal{O}(\log(1/\delta))$.

(2) The parameter $C(A)$ in \eqref{CA} is now replaced by $C(\bar{A}) = \sup_{t\in [0,T]} \|\e^{\bar{A} t}\|$, where
\[\|\e^{\bar{A} t}\| = \sup_{\|\bb{w}(0)\| \le 1}\|\e^{\bar{A} t}\bb{w}(0)\|. \]
Let $\bb{w}(t)=\e^{\bar{A} t}\bb{w}(0)$ with $\|\bb{w}(0)\| \le 1$. We can define $\tilde{\bb{w}}(t)$ and write it as the block structure in \eqref{tildewblock}, which gives
\[\frac{\d}{\d t} \bb{u}^{(l)}(t) = B(s_l)\bb{u}^{(l)}(t) = \begin{bmatrix}
A(s_l)  &  F(s_l) \\
O     &  O
\end{bmatrix}\bb{u}^{(l)}(t), \qquad l = 0,1,\cdots,N_s-1,\]
with the initial data satisfying
\[\sum_{l=0}^{N_s-1} \|\bb{u}^{(l)}(0)\|^2  = \|\bb{w}(0)\|^2 \le 1.\]
Let $\bb{u}^{(l)} = [\bb{x}^{(l)}; \bb{r}^{(l)}]$. One has $\frac{\d \bb{r}^{(l)}}{\d t} = \bb{0}$, hence $\bb{r}^{(l)} = \bb{r}^{(0)}$ is a constant vector. This implies
\[
\dfrac{\d \bb{x}^{(l)}(t)}{\d s} = A(s_l)\bb{x}^{(l)}(t) +  \tilde{\bb{b}} (s_l),  \qquad l = 0,1,\cdots, N_s-1
,\]
where the $i$-th entry of $\tilde{\bb{b}}(s_l)$ is
\[\tilde{b}_i(s_l) = r_i^{(0)} f_i (s_l), \qquad  f_i(s_l) = \frac{b_i (s_l)}{( (b_i^2)_{ave} + \varepsilon^2 )^{1/2}} . \]
The solution can be expressed as
\[\bb{x}^{(l)}(t) = \e^{A(s_l) t}\bb{x}^{(l)}(0) + \int_0^t \e^{A(s_l) \tau } \tilde{\bb{b}} (s_l) \d \tau, \qquad l = 0,1,\cdots, N_s-1,\]
which gives
\begin{align*}
\|\bb{x}^{(l)}(t)\|
& \le \|\e^{A(s_l) t}\bb{x}^{(l)}(0)\| + \int_0^t \| \e^{A(s_l) \tau } \tilde{\bb{b}} (s_l)\|  \d \tau \\
& \le \max_{t\in [0,T]} \| \e^{A(s_l) t } \| +  T \max_{t\in [0,T]} \| \e^{A(s_l) t } \| \|\bb{f} (s_l)\| \\
& =\max_{t\in [0,T]} \| \e^{A(s_l) t } \| ( 1 + T \|\bb{f} (s_l)\|) = C(A(s_l))( 1 + T \|\bb{f} (s_l)\|) ,
\end{align*}
where $C(A(s_l))$ is defined as \eqref{CA}.
As before, we have
\begin{align*}
\|\bb{w}(t)\|^2
& = \|\bb{w}(0)\|^2 + \sum_{l=0}^{N_s-1} ( \|\bb{x}^{(l)}(t)\|^2 - \|\bb{x}^{(l)}(0)\|^2)  \\
& \le \|\bb{w}(0)\|^2 - \sum_{l=0}^{N_s-1}  \|\bb{x}^{(l)}(0)\|^2  + \sum_{l=0}^{N_s-1}  (C(A(s_l))( 1 + T \|\bb{f} (s_l)\|))^2\\
& = \sum_{l=0}^{N_s-1}  \|\bb{r}^{(l)}(0)\|^2  + \sum_{l=0}^{N_s-1}  (C(A(s_l))( 1 + T \|\bb{f} (s_l)\|))^2 \\
& \le 1 + \sum_{l=0}^{N_s-1}  ( C(A(s_l))( 1 + T \|\bb{f} (s_l)\|))^2,
\end{align*}
yielding
\[C(\bar{A}) \le 1 + N_s^{1/2} \max_{0\le l \le N_s-1}C(A(s_l))( 1 + T \|\bb{f} (s_l)\|),\]
where $N_s = \mathcal{O}(\log (1/\delta))$.

(3) The sparsity of $\bar{A}$ is $N_s + s + 1$. Plugging these quantities in the time complexity given by Theorem \ref{thm:complexity}, we conclude that there exists a quantum algorithm that produces a state $\delta$-close to $\bb{x}(T)/\|\bb{x}(T)\|$, using
\[\mathcal{O}\Big( C(\bar{A}) \bar{s} T \|\bar{A}\| \cdot \mathrm{poly} \Big( \log \frac{C(\bar{A}) \bar{s} T \|\bar{A}\| \bar{g} \bar{\beta}   }{\delta}  \Big) \Big)\]
queries to $O_{\bar{A}}$ and $O_w$, where
\[\bar{\beta} = 1, \qquad C(\bar{A}) \le  1 + N_s^{1/2} \max_{l}C(A(s_l))( 1 + T \|\bb{f} (s_l)\|) ,\]
\[\bar{s} = N_s + s + 1, \qquad \|\bar{A}\| \le \frac{\pi N_s}{2} + \max_l \|A(s_l)\| + \max_{i,l} |f_i(s_l)|,\]
\[\bar{g} = \frac{\max_{t} \|\bb{w}(t)\|}{\|\bb{x}(t)\|}, \quad \max_{t\in [0,T]} \|\bb{w}(t)\| \le  N_s^{1/2} \Big( \|\bb{b}\|_{ave}^2+ 1 + \max_{0\le l \le N_s} \max_{t\in [0,T]} \|\bb{x}^{(l)}(t)\|^2 \Big)^{1/2}. \]
Since $a+b \le 2b \le ab$ for two numbers $b,a$ satisfying $b \ge a \ge 2$, the complexity simplifies to $\mathcal{O}( K_1 T \mathrm{poly}(\log (K_1 T g_1/\delta)))$, with
\begin{align*}
K_1
& = N_s^{1/2}(N_s+s) \Big( N_s + \max_l \|A(s_l)\| + \max_{i,l} |f_i(s_l)| \Big)  \max_{l} C(A(s_l))( 1 + T \|\bb{f} (s_l)\|)   \\
& \lesssim s N_s^{2.5} \Big( 1 + \max_l \|A(s_l)\| + \max_{i,l} |f_i(s_l)| \Big)  \max_l C(A(s_l))( 1 + T \|\bb{f} (s_l)\|)
\end{align*}
and
\[g_1 = N_s^{1/2} \frac{\max_{l,t} \|\bb{x}^{(l)}(t)\| + \|\bb{b}\|_{ave}+ 1 }{\|\bb{x}(t)\|}, \qquad N_s = \mathcal{O}(\log(1/\delta)). \]
The proof is finalized by incorporating terms related to $\log(1/\delta)$ within the $\text{poly}(\log)$ notation.
\end{proof}

\section{Conclusions}

In this paper, we have established a new complexity bound for the original quantum ODE solver introduced in \cite{BerryChilds2017ODE}. Our analysis parallels that of \cite{KroviODE}, providing comparable bounds on condition number and solution error through the introduction of a novel parameter outlined in \cite{KroviODE}. This reaffirms the inherent advantages of the BCOW approach in quantum differential equation solving. Additionally, we have extended the BCOW algorithm to handle time-dependent linear differential equations by transforming non-autonomous systems into higher-dimensional autonomous ones, a technique also adaptable within the framework of the Krovi algorithm.

\section*{Acknowledgements}

XJD is supported by the National Natural Science Foundation of China (No: 12071404), Young Elite Scientist Sponsorship Program by CAST (No: 2020QNRC001), the Science and Technology Innovation Program of Hunan Province (No: 2024RC3158). QLT is supported by Key Project of Scientific Research Project of Hunan Provincial Department of Education (No: 22A0136). Y Yang is  supported by the National Natural Science Foundation of China (No: 12261131501). YY is supported by the National Science Foundation for Young Scientists of China (No. 12301561), with additional partial support from NSFC grant No. 12341104.


\newcommand{\etalchar}[1]{$^{#1}$}

\end{document}